\documentclass[aps,pra,twocolumn,showpacs]{revtex4-1}
\usepackage{amsmath,amscd,amssymb,graphicx}
\usepackage{hyperref}
\newcommand{\ie}{\textit{i.e.}}
\begin{document}
\title{Experimental Detection of Qubit-Ququart Pseudo-Bound 
Entanglement using Three Nuclear Spins}
\author{Amandeep Singh}
\email{amandeepsingh@iisermohali.ac.in}
\affiliation{Department of Physical Sciences, Indian
Institute of Science Education \& 
Research Mohali, Sector 81 SAS Nagar, 
Manauli PO 140306 Punjab India.}
\author{Akanksha Gautam}
\email{akankshagautam@iisermohali.ac.in}
\affiliation{Department of Physical Sciences, Indian
Institute of Science Education \& 
Research Mohali, Sector 81 SAS Nagar, 
Manauli PO 140306 Punjab India.}
\author{Arvind}
\email{arvind@iisermohali.ac.in}
\affiliation{Department of Physical Sciences, Indian
Institute of Science Education \& 
Research Mohali, Sector 81 SAS Nagar, 
Manauli PO 140306 Punjab India.}
\author{Kavita Dorai}
\email{kavita@iisermohali.ac.in}
\affiliation{Department of Physical Sciences, Indian
Institute of Science Education \& 
Research Mohali, Sector 81 SAS Nagar, 
Manauli PO 140306 Punjab India.}
\begin{abstract}
In this work, we  experimentally created and characterized a
class of qubit-ququart PPT (positive under partial
transpose) entangled states using three nuclear spins on an
nuclear magnetic resonance (NMR) quantum information
processor.  Entanglement detection and characterization for
systems with a Hilbert space dimension  $ \geq 2 \otimes 3$
is nontrivial since there are states in such systems which
are both PPT as well as entangled.  The experimental
detection scheme that we devised for the detection of
qubit-ququart PPT entanglement was based on the measurement
of  three Pauli operators with high precision, and is a key
ingredient of the protocol in detecting entanglement.  The
family of PPT-entangled states considered in the current study are
incoherent mixtures of five pure states.  All the five
states were prepared with high fidelities and the resulting
PPT entangled states were prepared with mean fidelity $ \geq
$ 0.95.  The entanglement thus detected  was validated by
carrying out full quantum state tomography (QST).
\end{abstract} 
\maketitle 
\section{Introduction}
Quantum entanglement is one of the most counter-intuitive
phenomena encountered in the quantum
world and is at the heart of 
the quantum classical 
divide~\citep{schrodinger-Naturwissenschaften-35}.
In the development of quantum information science, it was
realized quite early that entanglement is a resource which
can be exploited to achieve a quantum advantage for
information processing and communication~\citep{nielsen-book-02}.
Specifically, entanglement  has applications in quantum
computing, quantum cryptography and key
distribution~\citep{bennett-jc-92,walborn-prl-06} and
quantum teleportation~\citep{riebe-nature-04}.
Theoretically
characterizing the entanglement of a given density operator
has been one of the core research areas in quantum
information
processing~\citep{horodecki-rmp-09}. While the
entanglement characterization of pure bipartite quantum
systems is well understood~\citep{peres-prl-96,horodecki-pla-96}, 
a complete  characterization of mixed
state entanglement remains
elusive~\citep{horodecki-book-01}.  For mixed states, for
two-qubit and qubit-qutrit systems, the entanglement can be
fully pinned down via the  well known PPT (positive under
partial transpose) criteria and such entanglement can also
be  distilled~\citep{bennett-pra-96} into the form of EPR
pairs~\citep{horodecki-prl-97}.  However, for
mixed states of quantum systems with Hilbert space
dimensions greater than $2 \otimes 3$ and for multiparty
situations, the problem of entanglement characterization is
still an open question~\citep{guhne-pr-09}.  Entanglement
exists in two fundamentally different
forms~\citep{horodecki-pla-97,horodecki-prl-98}: ``free''
entanglement which can be distilled into EPR pairs using
local operations and classical communications
(LOCC)~\citep{nielsen-prl-99} and ``bound'' entanglement
which cannot be distilled into EPR pairs via LOCC.

Even in situations where entanglement can be characterized
theoretically, the experimental detection of entanglement is
resource-intensive and remains a challenging
task~\cite{spengler-pra-12,dai-prl-14}.  
Several experimental efforts
in this direction have tried to reduce the resources
required to detect entanglement and have devised methods
based on entanglement witnesses and positive maps to
interrogate the presence of
entanglement~\cite{filgueiras-qip-12,yu-china-17,
singh-pra-16,singh-pra-18,singh-qip-18}.  A range of
experiments have been carried out to create and detect novel
entangled states~\citep{
bouwmeester-prl-99,
roos-sc-04,bourennane-prl-04,leibfried-nature-05,
haffner-nature-05,das-pra-15,dogra-pra-15,singh-pra-18-1}.

The term ``bound'' essentially implies that although
correlations were established during the state preparation,
they cannot brought into a ``free'' form in terms of EPR
pairs by a distillation process, and used wherever EPR
pairs can be used as a resource. PPT entangled states are a
prime example of bound entangled states and have been shown
to be useful to establish a secret
key~\citep{horodecki-prl-05}, in the conversion of pure
entangled states~\citep{ishizaka-prl-04} and for quantum
secure communication~\citep{zhou-prl-18}.  While the
existence of ``bound'' entangled states has been proved
beyond doubt, there are still only a few known classes of
such
states~\cite{horodecki-pla-97,smolin-pra-01,acin-prl-01,sengupta-pra-11}.
The problem of finding all such PPT entangled states is
still unsolved at the theoretical level.

Experimentally, bound entanglement has been created using
four-qubit polarization
states~\citep{amselem-natureP-09,kaneda-prl-12} and
entanglement unlocking of a four-qubit bound entangled state
was also demonstrated~\citep{lavoie-prl-10}.  Entanglement
was characterized in bit-flip and phase-flip lossless 
quantum channel and the experiments were able to
differentiate between free entangled, bound-entangled and
separable states~\citep{amselem-sr-13}.  Continuous variable
photonic bound-state entanglement has been created and
detected in various
experiments~\citep{dobek-lp-13,diGuglielmo-prl-11,steinhoff-pra-14}.
Two photon qutrit Bound-entangled states of two qutrits were
investigated utilizing orbital angular momentum degrees of
freedom~\citep{hiesmayr-njp-13}. In NMR, a three-qubit
system was used to prepare a three-parameter pseudo-bound
entangled state~\citep{suter-pra-10}.

In the present study, we experimentally create and
characterize a one-parameter family of qubit-ququart PPT
entangled states using three nuclear spins on an nuclear
magnetic resonance (NMR) quantum information
processor. There are a few
proposals to detect PPT entanglement in the class of states
introduced in Reference~\citep{horodecki-pla-97} by
exploring local sum uncertainties~\citep{zhao-pra-13} and by
measuring individual spin magnetisation along different
directions~\citep{akbari-ijqi-17}. We chose the proposal of
Reference~\citep{akbari-ijqi-17} to experimentally detect
PPT entanglement in states prepared on an NMR quantum
information processor.  The family of states considered in
the current study is an incoherent mixture of five pure
states and the relative strengths of the components  of 
the mixture is controlled by a real parameter. 
We experimentally prepared different such PPT-entangled states,
parameterized by a real parameter. These states represent
five points on the one-parameter family of states. Discrete
values of the real parameter were used which were uniformly
distributed over the range for which the current detection
protocol detects the entanglement.  In order to
experimentally detect entanglement in these states, three
Pauli operators need to be measured in each case. To measure
the required observables we utilized previously developed
schemes~\citep{singh-pra-16,singh-pra-18,singh-qip-18}
which unitarily map the desired state followed by  NMR
ensemble average measurements. In each case we also
performed full quantum state tomography
(QST)~\citep{leskowitz-pra-04,singh-pla-16} to verify the
success of the detection protocol as well as to establish
that the experimentally created states are indeed PPT
entangled.  Our work is important both in the context of
preparing and characterizing bound entangled states and in
devising new experimental schemes to detect PPT entangled
states which use much fewer resources than are required by
full quantum state tomography schemes.  It should be noted
here that we prepare the PPT entangled states using NMR in
the sense that the total density operator for the spin
ensemble always remains close to the maximally mixed state
and at any given instance we are dealing with
pseudo-entangled states~\citep{laflemme-ptsca-98}.

This article is organized as follows:~Sec.~\ref{theory} characterizes bound
entanglement in qubit-ququart systems and describes the theoretical aspects of
the detection scheme.  Sec.~\ref{NMR-Implementation} contains the main results
including the experimental NMR implementation of the PPT entanglement detection
scheme.  Sec.~\ref{concludingRemarks} contains a few concluding remarks. 
\section{Bound Entanglement in a Qubit-Ququart System}
\label{theory}
Consider a 3-qubit quantum system with an eight-dimensional
Hilbert space ${\mathcal H}={\mathcal H}_1\otimes{\mathcal
H}_2{}\otimes{\mathcal H}_3$, where ${\mathcal H}_i$
represent qubit Hilbert spaces.  If we choose to club the
last two qubits into a single system with a four-dimensional
Hilbert space ${\mathcal H}_q={\mathcal
H}_2{}\otimes{\mathcal H}_3$, the three-qubit system can
be reinterpreted as a qubit-ququart bipartite system with
Hilbert space ${\mathcal H}={\mathcal H}_1\otimes{\mathcal
H}_q$. 
Formally we can say that the four ququart basis vectors
$\vert e_i \rangle$ are mapped to the logical state vectors
of
the second and third qubits as $ \vert e_1 \rangle
\leftrightarrow \vert 00 \rangle ,\; \vert e_2 \rangle
\leftrightarrow \vert 01 \rangle ,\; \vert e_3 \rangle
\leftrightarrow \vert 10 \rangle$ and $\vert e_4 \rangle
\leftrightarrow \vert 11 \rangle $ in the computational
basis. 
With this understanding, we will freely use the three-qubit 
computational basis for this qubit-ququart system, where all along 
it is understood that the last two qubits form a ququart. 
For this system consider a family of   
PPT bound entangled states parameterized by a real
parameter  $ b \in (0,1)$
introduced by
Horodecki~\citep{horodecki-pla-97}.  
\begin{eqnarray}
\label{BE}
\sigma_b=\frac{7b}{7b+1}\sigma_{\rm insep}+
\frac{1}{7b+1}\vert \phi_b \rangle \langle \phi_b\vert
\end{eqnarray} 
with \begin{eqnarray}\label{BE-1} \sigma_{\rm
insep}&=&\frac{2}{7}\sum_{i=1}^3 \vert \psi_i\rangle\langle
\psi_i \vert+\frac{1}{7}\vert 011 \rangle\langle 011 \vert,
\nonumber\\ \vert \phi_b\rangle &=& \vert 1 \rangle \otimes
\frac{1}{\sqrt{2}}\left(\sqrt{1+b}\vert 00
\rangle+\sqrt{1-b}\vert 11 \rangle
\right),
\nonumber\\ \vert \psi_1 \rangle &=&
\frac{1}{\sqrt{2}}(\vert 000 \rangle + \vert 101
\rangle),\nonumber\\ \vert \psi_2 \rangle &=&
\frac{1}{\sqrt{2}}(\vert 001 \rangle + \vert 110
\rangle),\nonumber\\ \vert \psi_3 \rangle &=&
\frac{1}{\sqrt{2}}(\vert 010 \rangle + \vert 111 \rangle)
\end{eqnarray}
 It has been shown in~\citep{horodecki-pla-97} that the
states in the family $\sigma_b$  defined above  are entangled
for $ 0<b<1 $ and is separable in the limiting cases $ b=0
\;\; \rm or \;\; 1 $. One can explicitly write the density
operator for the mixed PPT entangled states defined in
Equation~(\ref{BE}) in the computational basis as 
\begin{equation}
\label{BE-mtrx}
\sigma_b=\frac{1}{1+7b}\left[ \begin{array}{cccccccc}
b&0&0&0&0&b&0&0\\ 0&b&0&0&0&0&b&0\\ 0&0&b&0&0&0&0&b\\
0&0&0&b&0&0&0&0\\
0&0&0&0&\frac{(1+b)}{2}&0&0&{\frac{\sqrt{1-b^2}}{2}}\\
b&0&0&0&0&b&0&0\\ 0&b&0&0&0&0&b&0\\
0&0&b&0&{\frac{\sqrt{1-b^2}}{2}}&0&0&\frac{(1+b)}{2}\\
\end{array} \right] 
\end{equation} 
It is interesting to observe that for $ b=0 $, this  family of
states reduce to a  separable state in 2$
\otimes $4 dimensions while it is still entangled in the 
three-qubit space and the entanglement  is restricted to the 
two qubits forming the ququart.

\begin{figure}[t]
\includegraphics[angle=0,scale=1]{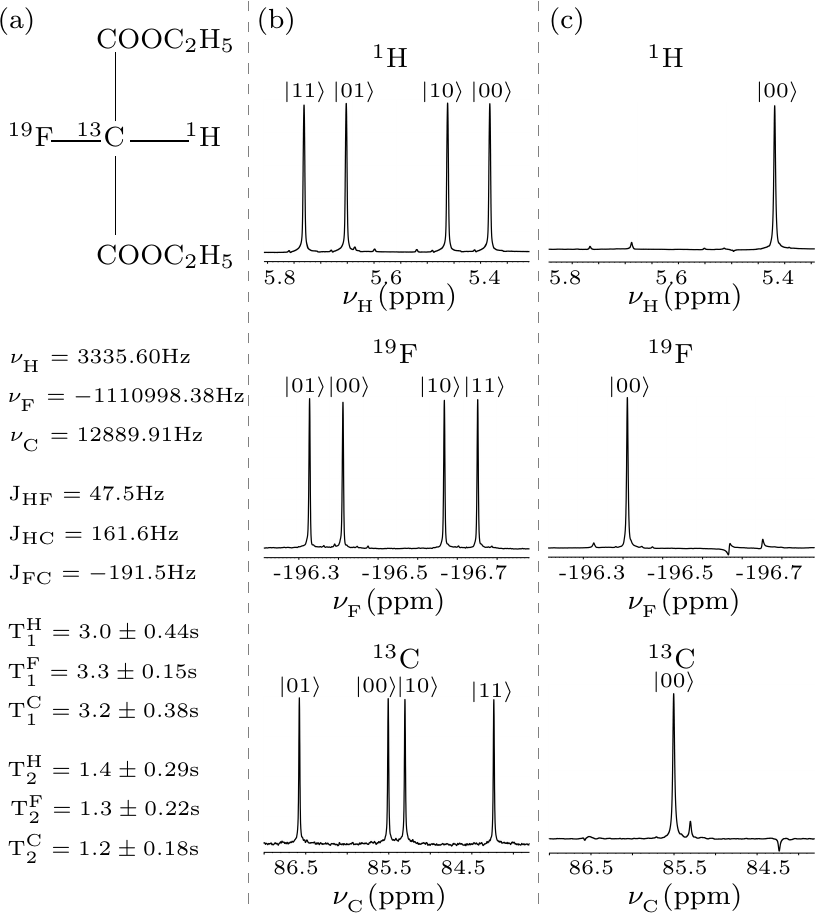}
\caption{(a) $^{13}$C-labeled diethylfluoromalonate molecule
and NMR parameters. NMR spectra of (b) thermal equilibrium
state and (c) $ \vert 000 \rangle $ pseudo pure state. Each
transition is labeled with the logical states using the
three-qubit basis}.
\label{molecule} 
\end{figure}

Having defined the family of PPT entangled  states in
Equation(\ref{BE-mtrx}) parameterized by `$ b $', we now
describe the method that we use to experimentally detect
their entanglement  using a protocol proposed in
Reference~\citep{akbari-ijqi-17}.  Although the family of
states (\ref{BE-mtrx}) is PPT entangled in $2 \otimes
4$-dimensional Hilbert space, it is useful to exploit the
underlying three-qubit structure.  For the detection
protocol, we define three observables $B_i$, with $i=1,2,3$
(here $B_1$ acting on the qubit space and $B_2$ and $B_3$
act in the state  space of qubits 2 and 3 forming the
ququart)
\begin{equation}\label{BE-obs}
B_1=\mathbb{I}_2 \otimes \sigma_x \otimes \sigma_x,  \;\;
B_2=\mathbb{I}_2 \otimes \sigma_y \otimes \sigma_y,  \;\;
B_3=\sigma_z \otimes \sigma_z \otimes \sigma_z
\end{equation} where $ \sigma_{x,y,z} $ are the
Pauli operators and $ \mathbb{I}_2 $ is the 2$ \times $2
identity operator. Although the observables $B_j$ defined
above are written in the three qubit notation, they are bona
fide observables of the qubit-ququart system. The main
result of Reference~\citep{akbari-ijqi-17} is that any
three-qubit separable state, $ \rho_s $, obeys the four
inequalities given by 
\begin{equation}\label{BE-inequality} \vert \langle B_1
\rangle_{\rho_s} \pm \langle B_2 \rangle_{\rho_s} \pm
\langle B_3 \rangle_{\rho_s} \vert \leq 1 
\end{equation}
Therefore, if a states violates even one of the
four inequalities given in Eq.~(\ref{BE-inequality}),
it has to be entangled.
It was shown numerically in~\cite{akbari-ijqi-17} that the
inequalities defined in Equation~(\ref{BE-inequality}) can
be used to detect the entanglement present in the states
$\sigma_b$ defined in Eq.~(\ref{BE-mtrx}) for  $
0<b<\frac{1}{\sqrt{17}}$.  Hence we are able to detect the
entanglement of this family of  PPT entangled in $2 \otimes
4$ dimensions.
\begin{figure}[t]
\includegraphics[angle=0,scale=1]{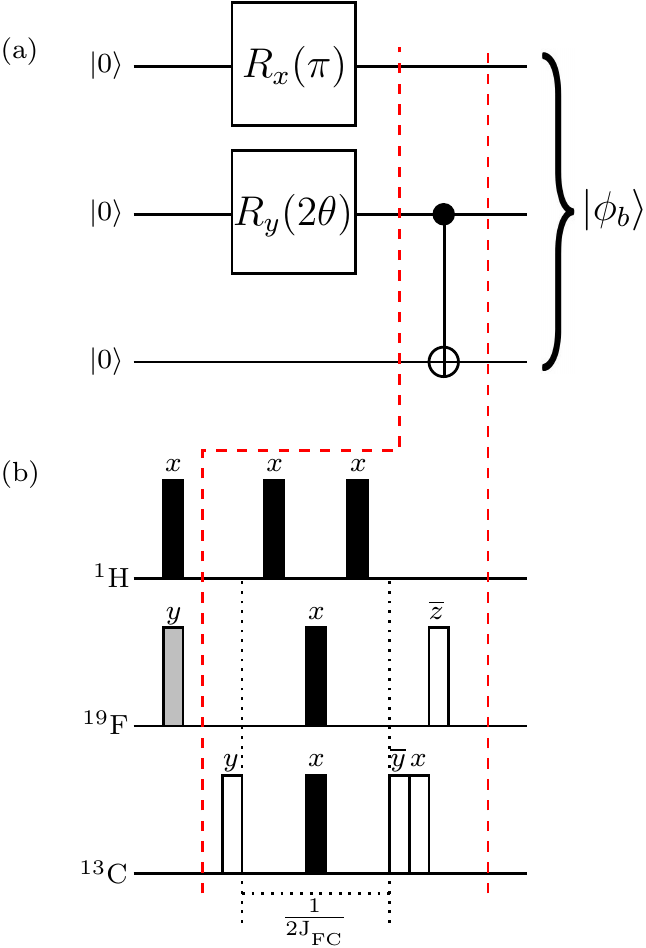}
\caption{(a) Quantum circuit to prepare  $ \vert \phi_b
\rangle $ from $ \vert 000 \rangle$ pseudopure state. (b)
NMR pulse sequence for quantum circuit given in (a).  Blank
rectangles represent $ \frac{\pi}{2} $ RF pulses, while
black rectangles represent $ \pi $ spin-selective rotations.
The gray rectangle represents a rotation through $
\theta=\;$Cos$^{-1}\sqrt{1+b}/\sqrt{2} $.  The phase of
each RF pulse is written above the respective pulse. A bar
over a phase implies negative phase, while the free
evolution time interval is given by $ (\rm
2J^{}_{FC})^{-1}$.}
\label{ckt} 
\end{figure}
\section{Experimental Detection of 2$ \otimes $4 Bound
Entanglement} 
\label{NMR-Implementation}
We now proceed toward experimentally preparing several
different states from the family of states given by
Eq.~(\ref{BE-mtrx}), detect them by measuring the
observables defined in Eq.~(\ref{BE-obs}) and check if we
observe a violation of the inequalities defined
in Eq.~(\ref{BE-inequality}). 
\begin{figure}[t]
\begin{center}
\includegraphics[angle=0,scale=1]{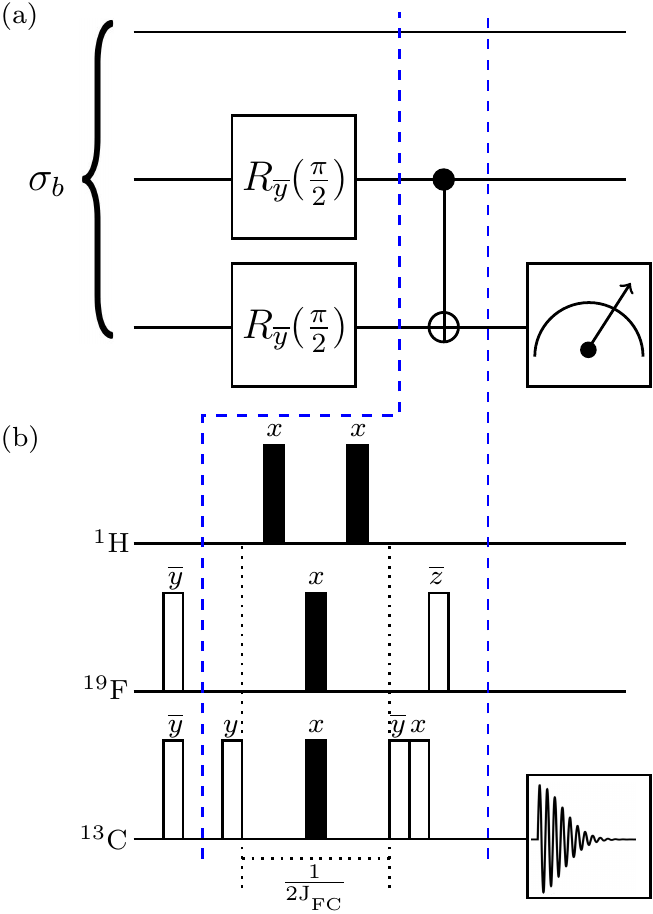}
\end{center}
\caption{(a) Quantum circuit to map $ \sigma_b $ to the
state $ \sigma^{\prime}_b $ such that $ \langle  B_1 \rangle
_{\sigma_b}= \langle I_{3z} \rangle _{\sigma_b^{\prime}} $.
(b) NMR pulse sequence to achieve the quantum circuit in
(a). The unfilled rectangles denote $ \frac{\pi}{2} $ RF
pulses, while the filled rectangles represent $\pi$
spin-selective RF pulses.  The phase of each RF pulse is
written above the respective pulse. A bar over a phase
implies negative phase and the free evolution time interval
is given by $ \rm \frac{1}{2J^{}_{FC}} $.}
\label{mapping_ckt} 
\end{figure}

In order to prepare the PPT entangled family of states in $2
\otimes 4$ dimensions using three qubits, we chose three
spin-1/2 nuclei ($^1$H, $^{19}$F and $^{13}$C)  to
encode the three qubits in a $^{13}$C-labeled sample of
diethylfluoromalonate dissolved in acetone-D6.  The
weak-field free evolution NMR Hamiltonian 
for the system is given by~\citep{ernst-book-90}
\begin{equation}\label{3Q-Hamiltonian}
 \mathcal{H}=-\sum_{i=1}^{3} \nu_i I_{iz} + \sum_{i>j,i=1}^{3}
J_{ij}I_{iz}I_{jz}
\end{equation}
where the indices $i,j=1,2\; \rm or \;3$ label the qubit, $
\nu_i $ are the respective chemical shifts, $ I_{iz} $ is
the Pauli $ z $ spin angular momentum operator of the $
i^{\rm th} $ spin and $ J_{ij} $ is the scalar spin-spin
coupling strength.  
The ensemble was first
initialized in the pseudopure state (PPS) $\vert 000
\rangle$ using the spatial
averaging method 
with the
PPS density operator being given 
by~\cite{cory-physD-98} 
\begin{equation}\label{PPS}
\rho_{000}=\frac{(1-\epsilon)}{2^3}\mathbb{I}_8 +\epsilon
\vert 000 \rangle \langle 000 \vert
\end{equation}
where $ \epsilon $ ($\sim\ 10^{-5}$) is the thermal spin
magnetization given by the Boltzmann factor ($ \rm \mu
B/k_BT $) of the spin ensemble placed in a static magnetic
field B at temperature T and $ \mathbb{I}_8 $ is the
8$\times$8 identity operator.  The NMR experimental
parameters 
as well as the spectra of
the thermal and PPS states are shown in Fig.(\ref{molecule}).
Each transition in the NMR spectra is labeled with the
logical state of the passive qubits. We experimentally
prepared the PPS with a fidelity of 0.98$ \pm $0.01. 
The state
fidelity was computed using the fidelity measure
\citep{uhlmann-rpmp-76,jozsa-jmo-94} 

\begin{equation}\label{fidelity}
\rm F=\left[Tr\left( \sqrt{\sqrt{\rho_{{\rm th}}}\rho_{{\rm
ex}} \sqrt{\rho_{{\rm th}}}}\right)\right]^2,
\end{equation}
where $\rho_{\rm th}$ and $\rho_{\rm ex}$ represent
theoretically expected and experimentally prepared density
operators, respectively and F is normalized in the sense
that $ \rm F \rightarrow 1 $ as $\rho_{\rm ex} \rightarrow
\rho_{\rm th}$.

\begin{figure}[t]
\includegraphics[angle=0,scale=1]{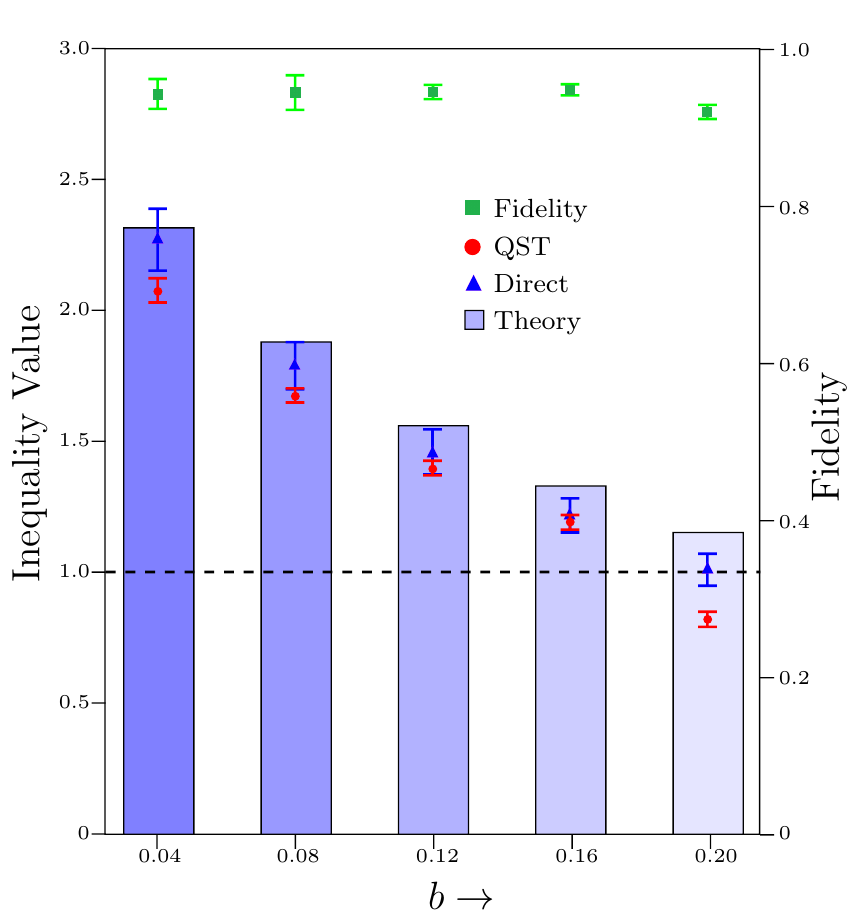}
\caption{(Color online) Bars represent theoretically
expected values, red circles are the values obtained via QST
and blue triangles are the direct experimental values for
the inequality appearing in Eq.(\ref{BE-obs}). Green squares
are the mean experimental fidelities. Horizontal black
dashed line is the reference line for states in
Eq.(\ref{BE}) violating inequality of Eq.(\ref{BE-obs}).}
\label{resultplot} 
\end{figure}

The experimentally prepared bound entangled states in the
current study were directly detected using the protocol
discussed in Sec.~\ref{theory} and full QST
\citep{leskowitz-pra-04} was also performed in each case to
verify the results. QST utilized a set of seven preparatory
pulses $\left\lbrace III, XXX, IIY, XYX, YII, XXY, IYY
\right\rbrace$, where $ I $ implies `no-operation' and $
X(Y) $ is a local $ \frac{\pi}{2} $ unitary rotation with
phase $ x(y) $. In NMR, such local unitary rotations are
achieved using highly accurate and calibrated spin-selective
radiofrequency (RF) pulses applied transverse to the static
magnetic field. Experiments were performed on a Bruker
Avance-III 600 MHz FT-NMR spectrometer equipped with QXI
probe at room temperature ($\sim$\; 20$^{\circ}$C). Three
dedicated channels for $ ^1 $H, $ ^{19} $F and $ ^{13} $C
nuclei were employed having $ \frac{\pi}{2} $ RF pulse
durations of 9.33 $\mu $s, 22.55 $\mu $s and 15.90 $\mu$s at
the power levels of 18.14 W, 42.27 W and 179.47 W
respectively.

The next step was to prepare the PPT entangled family of
states given in Eq.~(\ref{BE-mtrx}) (each with a fixed value of
the parameter $b$) and to achieve this we utilized the
method of temporal averaging~\citep{cory-physD-98}. 
The family of states $\sigma_b$ is an incoherent mixture of
several pure states as given in Equation(\ref{BE-1}), and the
quantum circuit to prepare one such nontrivial state ($
\vert \phi_b \rangle $) is given in Fig.\ref{ckt}(a), where
$ R_x(\pi) $ represents a local unitary rotation through an
angle $ \pi $ with a phase $x$.  After experimentally
preparing the state, one can measure the desired observable
in Eq.~(\ref{BE-obs}), by mapping the state
onto the Pauli basis operators.
 The quantum
circuit to achieve this is shown in
Fig.\ref{mapping_ckt}(a), and this circuit maps the state $ \sigma_b
\rightarrow \sigma_b^{\prime}$  such that $ \langle B_1
\rangle_{\sigma_b} = \langle I_{3z}
\rangle_{\sigma_b^{\prime}} $.  The motivation for such
a mapping~\citep{singh-pra-16,singh-pra-18,singh-qip-18}
relies on the fact that in an NMR scenario, the expectation
value  $ \langle I_{z} \rangle $, can be readily measured
\citep{ernst-book-90}.
The crux of the temporal averaging technique relies on the
fact that 
the five states composing the PPT entangled state are generated
via five different experiments. The states of these experiments
are then added with appropriate probabilities to achieve the
desired PPT entangled state.
\begin{figure}[t]
\includegraphics[angle=0,scale=1]{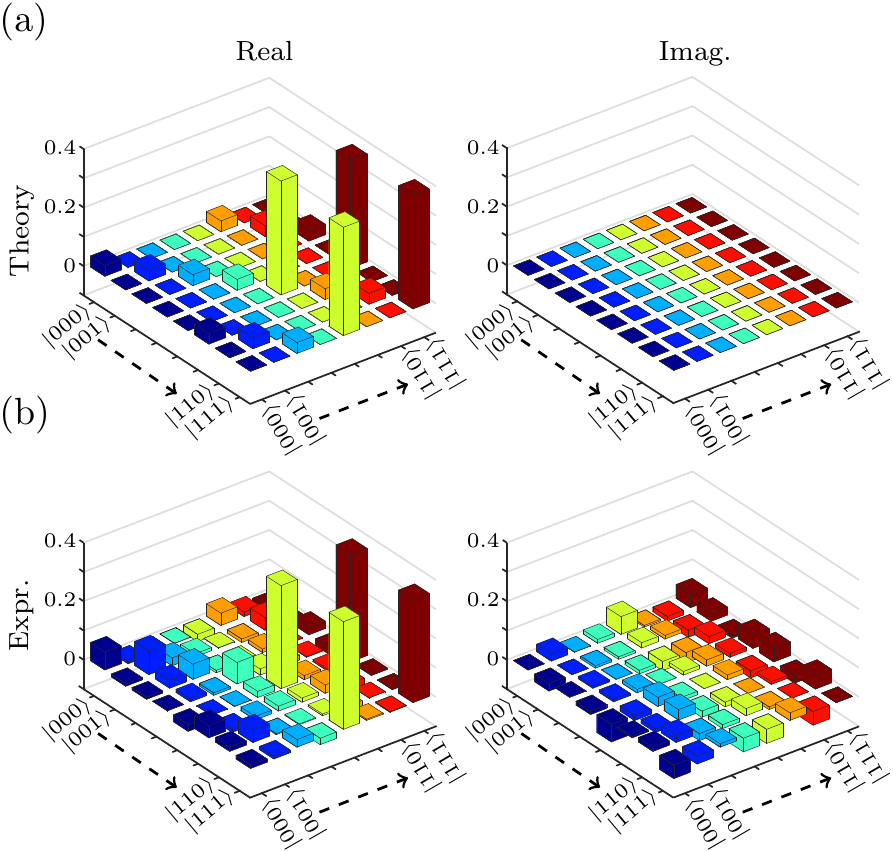}
\caption{Real and imaginary parts of the tomograph of the
(a) theoretically expected and (b) experimentally
reconstructed density operator for PPT entangled state with
$b=0.04$ and state fidelity F=0.968.}
\label{tomo} 
\end{figure}

All the
five states, appearing in Eq.(\ref{BE-1}), \ie\; $ \vert
\phi_b \rangle $, $ \vert \psi_1 \rangle $, $ \vert \psi_2
\rangle $, $ \vert \psi_3 \rangle $ and the separable PPS
state $ \vert 011 \rangle $ were experimentally prepared
with state fidelities $ \geq $ 0.96. It is worthwhile to note
here that $ \vert \phi_b \rangle  $ is a generalized
biseparable state while $ \vert \psi_1 \rangle $ and $ \vert
\psi_3 \rangle $ are LOCC equivalent biseparable states with
maximal entanglement between the first and third qubits and
$ \vert \psi_2 \rangle $ is a state belonging to the GHZ class.
For the experimental demonstration of the detection protocol
discussed in Sec.\ref{theory} we chose the values $
b=0.04,\; 0.08,\; 0.12,\; 0.16 \; \rm and \; 0.20 $  and 
prepared five different PPT entangled states.  The quantum
circuit as well as the NMR pulse sequence to prepare $ \vert
\phi_b \rangle  $ is shown in Fig.(\ref{ckt}). Other states
in Eq.(\ref{BE-1}) have similar circuits as well as pulse
sequences and are not shown here.  The tomograph for one
such experimentally prepared PPT entangled state, with $
b=0.04 $ and fidelity $ \rm F=0.968 $, is shown in
Fig.(\ref{tomo}). In order to measure the expectation values
of the observables appearing in Eq.(\ref{BE-obs}) we
utilized the procedure developed in our earlier
work~\citep{singh-pra-16,singh-pra-18}.  The idea is to
unitarily map the state $ \sigma_b $ to a state say $
\sigma_b^{\prime} $, such that $ \langle  \mathbb{O} \rangle
_{\sigma_b}= \langle  I_{iz} \rangle _{\sigma_b^{\prime}} $
where $ \mathbb{O} $ is one of the observables
to be measured in the state $ \sigma_b $.  This is achieved by
measuring $ I_{iz} $ on $ \sigma_b^{\prime} $. 
As an example, one can find the expectation value $ \langle B_1
\rangle_{\sigma_b} $ using the quantum circuit given in
Fig.\ref{mapping_ckt}(a)  and the NMR
pulse sequence given in Fig.\ref{mapping_ckt}(b) 
is implemented, followed by 
a measurement of the spin magnetization of the
third qubit.  Such a normalized magnetization of a qubit in
the mapped state is indeed proportional to the expectation
value of the $z$-spin angular momentum of the
qubit~\citep{ernst-book-90}.
\begin{table} [h]
\caption{\label{result table}
Experimentally measured values of 
the inequality in Eq.~(\ref{BE-inequality}) 
showing maximum violation for five different PPT entangled states.}
\begin{ruledtabular}
\begin{tabular}{c | c | c | c | c }

Obs. $\rightarrow$ &  &\multicolumn{3}{c}{Inequality value from: } \\

State(F) $\downarrow$ & $b$ & Theory & QST & Experimental\\
\colrule
$\sigma_{b_1}$(0.946$\pm$0.019) & 0.04 & 2.311 &
2.061$\pm$0.046 & 2.269$\pm$0.118 \\
$\sigma_{b_2}$(0.947$\pm$0.022) & 0.08 & 1.876 &
1.660$\pm$0.027  & 1.784$\pm$0.090 \\
$\sigma_{b_3}$(0.949$\pm$0.009) & 0.12 & 1.557 &
1.382$\pm$0.028  & 1.451$\pm$0.086 \\
$\sigma_{b_4}$(0.953$\pm$0.007) & 0.16 & 1.327 &
1.179$\pm$0.028  & 1.213$\pm$0.065 \\
$\sigma_{b_5}$(0.925$\pm$0.009) & 0.20 & 1.150 &
0.807$\pm$0.029 & 1.007$\pm$0.061 \\
\end{tabular}
\end{ruledtabular}
\end{table}
Experimentally measured values of the
inequality given in Eq.~\ref{BE-inequality} with maximum violation are
reported in Table-\ref{result table}. For all five states
with different $b$ values, full QST was also performed and
the observables \ref{BE-obs} were analytically computed from
the reconstructed density operators.  All the experimental
results, tabulated in the Table-\ref{result table}, are
plotted in Fig.~(\ref{resultplot}). Blue bars represent the
theoretically expected values, red circles are the values
obtained via full QST and blue triangles are the direct
experimental values for the inequality appearing in
Eq.(\ref{BE-obs}). Green squares are the mean experimental
fidelities. Horizontal black dashed line is the reference
line for states in Eq.(\ref{BE}) violating inequality of
Eq.(\ref{BE-obs}).  All the experiments were performed
several times to ensure the reproducibility of the
experimental results as well as to estimate the errors
reported in Table~\ref{result table}. It was observed that
the experimental values, within experimental error limits,
agree well with theoretically expected values and validate
the success of the detection protocol in
identifying the PPT entangled family of
states. The direct QST based measurements of
the state also validate our experimental results.
\section{Concluding Remarks}
\label{concludingRemarks}
The characterization of bound entangled states is
useful since it sheds light on the relation between
intrinsically quantum phenomena such as entanglement
and nonlocality.
The detection of bound entangled states is theoretically
a hard task and there are as yet no simple methods to
characterize all such states for arbitrary composite
quantum systems. 
The structure of PPT entangled states is rather complicated
and does not easily lead to a simple parametrization in terms
of a noise parameter. In this work,
we have reported the experimental creation of a family of
PPT entangled states of a qubit-ququart system  and the
implementation of a detection protocol involving local
measurements to detect their bound entanglement.  Five
different states which were parameterized by a real
parameter `$ b $', were experimentally prepared (with state
fidelities $ \geq $ 0.95) to represent the PPT entangled
family of states.  All the experiments were repeated several
times to ensure the reproducibility of the experimental
results and error estimation. In each case we observed that
the detection protocol successfully detected the PPT
entanglement of the state in question within experimental
error limits.  The results were further substantiated via
full QST for each prepared state.  It would be interesting
to create the PPT entangled family of states using different
pseudopure creation techniques and in higher dimensions, and
these directions will be taken up in future work.
\begin{acknowledgments}
All the experiments were performed on a Bruker Avance-III
600 MHz FT-NMR spectrometer at the NMR Research Facility of
IISER Mohali. Arvind acknowledges funding from DST India
under Grant No. EMR/2014/000297. K.D. acknowledges funding
from DST India under Grant No. EMR/2015/000556.
\end{acknowledgments}

%
\end{document}